\documentclass[fleqn,twoside]{article}
\usepackage{espcrc2}

\usepackage{graphicx}

\usepackage{floatflt}

\newcommand{\snn}{\sqrt{s_{NN}}}
\newcommand{\seff}{\s_{\rm eff}}
\newcommand{\s}{\sqrt{s}}
\newcommand{\pp}{pp}
\newcommand{\pbarp}{\overline{p}p}

\newcommand{\epem}{e^+e^-}

\newcommand{\nch}{N_{ch}}
\newcommand{\np}{N_{part}}

\newcommand{\ntot}{\langle\nch\rangle}
\newcommand{\avenp}{\langle\np\rangle}

\newcommand{\nc}{N_{coll}}

\newcommand{\halfnp}{\langle\np/2\rangle}
\newcommand{\etap}{\eta^{\prime}}

\newcommand{\dndeta}{d\nch/d\eta}
\newcommand{\dndetap}{d\nch/d\etap}

\newcommand{\dndetanp}{\dndeta / \halfnp}
\newcommand{\dndetapnp}{\dndetap / \halfnp}

\newcommand{\ratio}{\ntot/\halfnp}

\newcommand{\nubar}{\overline{\nu}}
\newcommand{\yb}{y_{\rm beam}}

\title{Universal Behavior of Charged Particle Production 
in Heavy Ion Collisions at RHIC Energies}

\author{Peter A. Steinberg$^{2}$ 
\thanks{Current address: Physics Department, University of Cape Town, South Africa}
for the PHOBOS collaboration\\
\vspace*{1mm}
{\footnotesize
B.B.Back$^1$,
M.D.Baker$^2$,
D.S.Barton$^2$,
R.R.Betts$^6$,
M.Ballintijn$^4$,
A.A.Bickley$^7$,
R.Bindel$^7$,
A.Budzanowski$^3$,
W.Busza$^4$,
A.Carroll$^2$,
M.P.Decowski$^4$,
E.Garc\'{i}a$^6$,
N.George$^{1,2}$,
K.Gulbrandsen$^4$,
S.Gushue$^2$,
C.Halliwell$^6$,
J.Hamblen$^8$,
G.A.Heintzelman$^2$,
C.Henderson$^4$,
D.J.Hofman$^6$,
R.S.Hollis$^6$,
R.Ho\l y\'{n}ski$^3$,
B.Holzman$^2$,
A.Iordanova$^6$,
E.Johnson$^8$,
J.L.Kane$^4$,
J.Katzy$^{4,6}$,
N.Khan$^8$,
W.Kucewicz$^6$,
P.Kulinich$^4$,
C.M.Kuo$^5$,
W.T.Lin$^5$,
S.Manly$^8$,
D.McLeod$^6$,
J.Micha\l owski$^3$,
A.C.Mignerey$^7$,
R.Nouicer$^6$,
A.Olszewski$^3$,
R.Pak$^2$,
I.C.Park$^8$,
H.Pernegger$^4$,
C.Reed$^4$,
L.P.Remsberg$^2$,
M.Reuter$^6$,
C.Roland$^4$,
G.Roland$^4$,
L.Rosenberg$^4$,
J.Sagerer$^6$,
P.Sarin$^4$,
P.Sawicki$^3$,
W.Skulski$^8$,
S.G.Steadman$^4$,
P.Steinberg$^2$,
G.S.F.Stephans$^4$,
M.Stodulski$^3$,
A.Sukhanov$^2$,
J.-L.Tang$^5$,
R.Teng$^8$,
A.Trzupek$^3$,
C.Vale$^4$,
G.J.van~Nieuwenhuizen$^4$,
R.Verdier$^4$,
B.Wadsworth$^4$,
F.L.H.Wolfs$^8$,
B.Wosiek$^3$,
K.Wo\'{z}niak$^3$,
A.H.Wuosmaa$^1$,
B.Wys\l ouch$^4$\\
\vspace{3mm}
\small
$^1$~Argonne National Laboratory, Argonne, IL 60439-4843, USA
$^2$~Brookhaven National Laboratory, Upton, NY 11973-5000, USA
$^3$~Institute of Nuclear Physics, Krak\'{o}w, Poland
$^4$~Massachusetts Institute of Technology, Cambridge, MA 02139-4307, USA
$^5$~National Central University, Chung-Li, Taiwan
$^6$~University of Illinois at Chicago, Chicago, IL 60607-7059, USA
$^7$~University of Maryland, College Park, MD 20742, USA
$^8$~University of Rochester, Rochester, NY 14627, USA

}
}

\begin{document}


\begin{abstract}
The PHOBOS experiment at RHIC has measured the multiplicity of
primary charged particles as a function of centrality and 
pseudorapidity in Au+Au collisions at $\snn = $ 19.6, 130 and 200 GeV.  
Two kinds of universal behavior are observed in charged particle production
in heavy ion collisions.
The first is that forward particle production, over a range of
energies, follows a universal limiting curve with a non-trivial
centrality dependence.
The second arises from comparisons with $\pp/\pbarp$ and $\epem$ data.  
$\ratio$ in nuclear collisions at high energy scales
with $\s$ in a similar way as $\nch$ in $\epem$ collisions and
has a very weak centrality dependence.
This feature may be related to a reduction in the leading
particle effect due to the multiple collisions suffered per participant
in heavy ion collisions.

\vspace*{-5mm}
\end{abstract}
\maketitle

\section{Introduction}
The PHOBOS experiment has measured $\dndeta$ and the
average multiplicity of charged particles $\ntot$ produced in heavy ion
collisions for center of mass energies in the nucleon-nucleon
center of mass system, $\snn$, of 19.6, 130 and 200 GeV.
The data is also binned 
as a function of event centrality (impact parameter) characterized
by the number of participating nucleons, $\np$, 
allowing comparisons to elementary systems, like 
$\pp/\pbarp$ and $\epem\rightarrow{\rm hadrons}$.

The PHOBOS multiplicity detector consists of several arrays of silicon
detectors which cover nearly the full solid angle for collision
events.
The event centrality is characterized by the multiplicity of charged
particles measured by two sets of 16 paddle counters covering $3<|\eta|<4.5$.
The methods used for measuring the multiplicity of charged particles 
as well as for extracting $\np$ 
has been described in more detail in Ref. \cite{phobos_limfrag}.

\section{Limiting Behavior in Pseudorapidity Distributions}

\begin{figure}[h]
\begin{center}
\includegraphics[width=8cm]{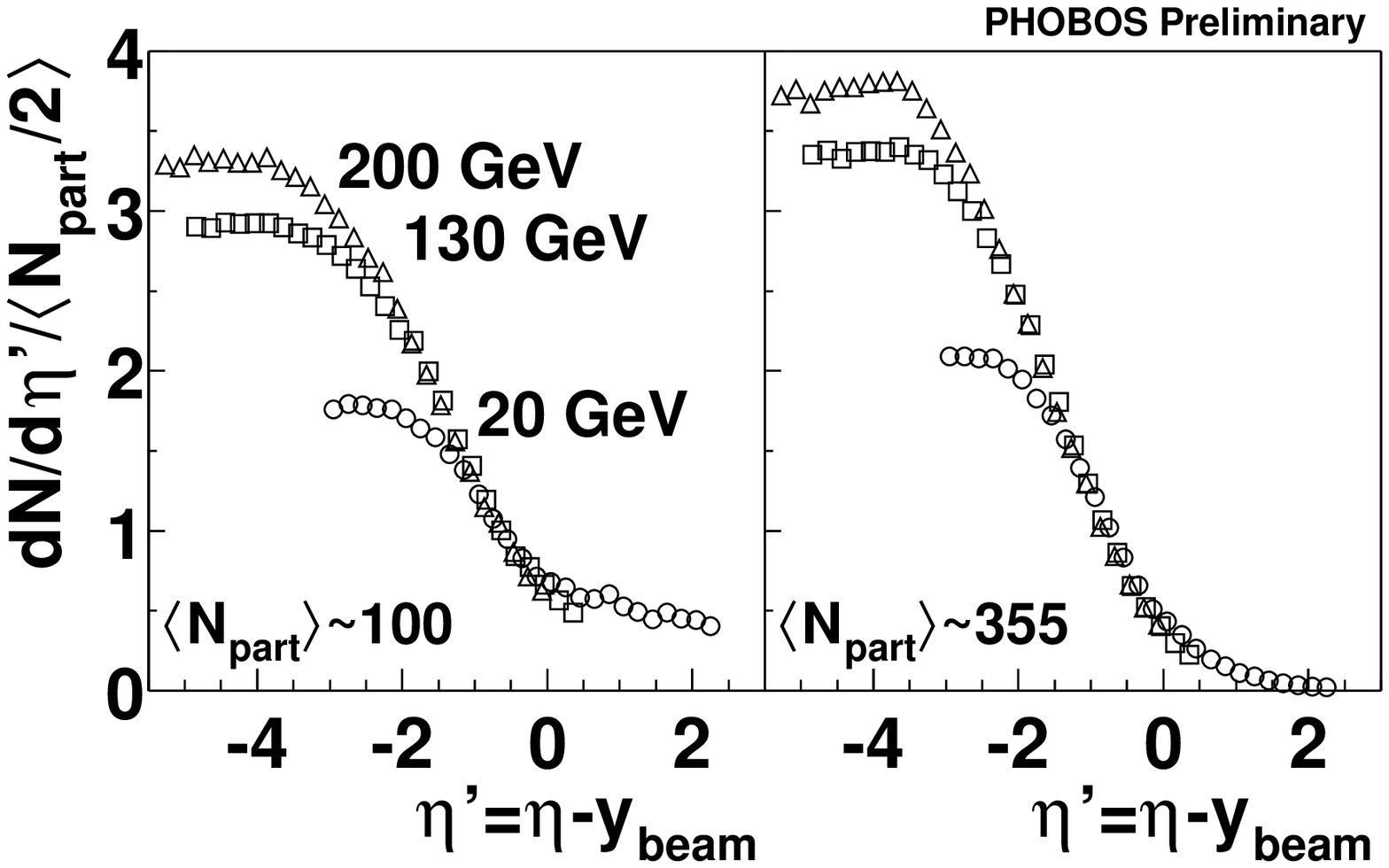}
\caption{$dN/d\etap/\halfnp$ for peripheral ($\avenp \sim 100$) 
and central ($\avenp \sim 355$) events at three RHIC energies.
\label{fig:limfrag}}
\end{center}
\vspace*{-.7cm}
\end{figure}

Fig. \ref{fig:limfrag} shows $\dndetapnp$ ($\etap = \eta-\yb$) 
measured at three different
RHIC energies for peripheral ($\avenp \sim 100$) and central events 
($\avenp \sim 355$),
in the left and right panels, respectively.
These show a clear ``limiting
behavior'' in the fragmentation region.
That is, the distributions are independent of beam energy in a substantial
range in $\etap$.
As the beam energy increases, $dN/d\etap$ follows the universal trend
until it reaches 85-90\% of it's maximum value at midrapidity, at which
point it stops following the trend.
Similar behavior has been observed in elementary collisions as well,
both in $\pbarp$ collisions \cite{ua5} 
and in $\epem$ collisions over a large range of
energies \cite{delphi}.

\vspace*{-.5cm}

The limiting curve constrains the energy dependence of 
the charged particle multiplicity.
It also
varies with centrality in such a way that the increases seen at low $\etap$ 
(which is midrapidity in $\eta$) as $\np$ increases, are 
accompanied by decreases near 
$\etap\sim 0$ (forward rapidities), as seen in Fig. \ref{fig:limfrag}.
It is not clear why this behavior occurs, e.g. whether it is from
energy conservation or a true long-range correlation.

\section{Comparison with Elementary Systems}

\begin{figure}[h]
\begin{center}
\includegraphics[height=70mm]{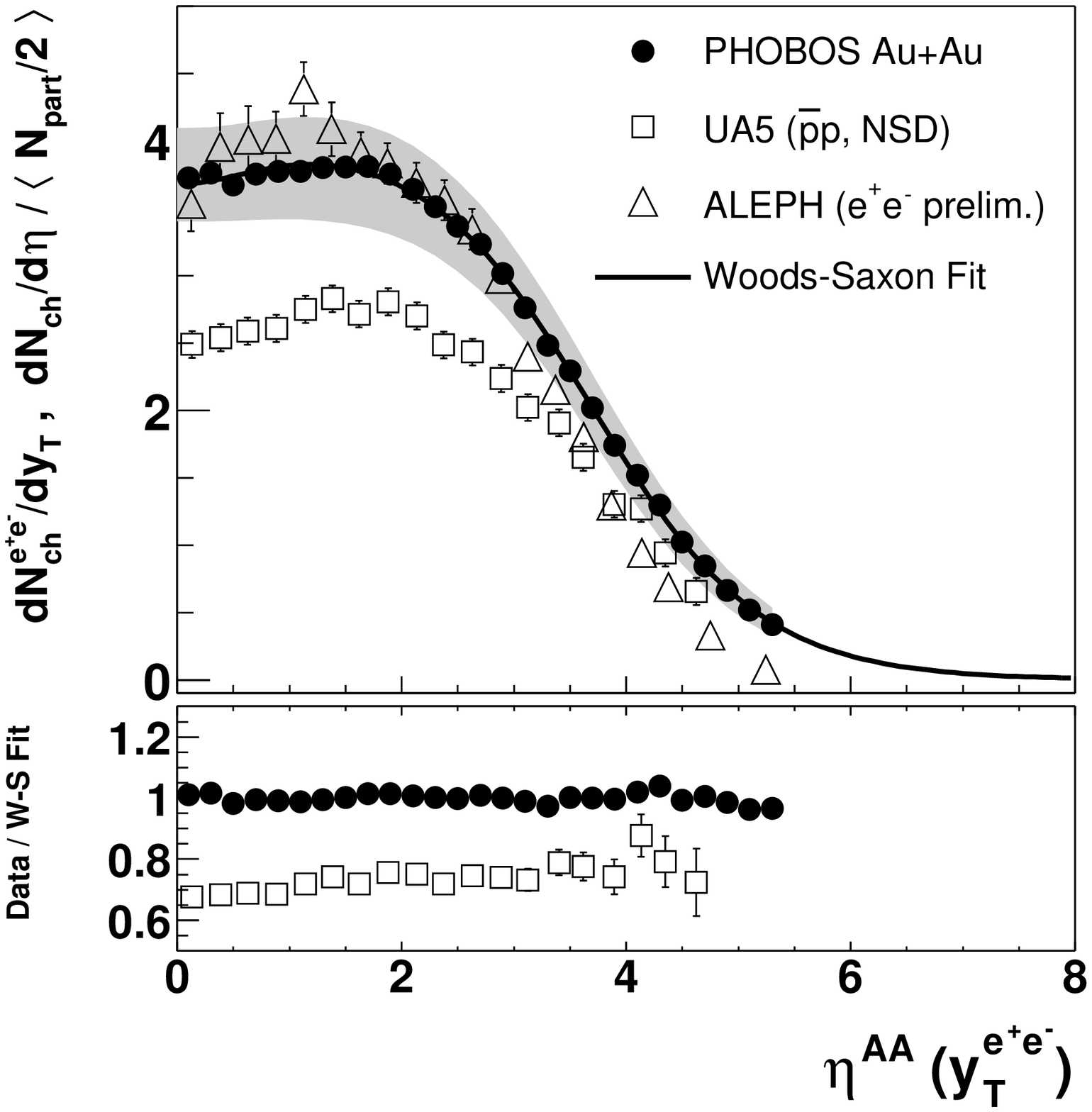}
\caption{
Top: $\dndetanp$ for central Au+Au collisions at $\snn=$ 200 GeV
compared with $\pbarp$ and $\epem$ data.
Bottom: Au+Au and $\pbarp$ data divided by a fit to the Au+Au data.
\label{fig:AA_ee_pp}
}
\end{center}
\vspace*{-.7cm}
\end{figure}

The angular distributions of charged particles for
different strongly-interacting systems are
shown in Fig. \ref{fig:AA_ee_pp}, where central Au+Au (divided by 
$\halfnp$), $\pbarp$ \cite{ua5} and $\epem$ \cite{ALEPH}
data are compared, all at $\s=200$ GeV.  In this figure, 
the Au+Au and $\pbarp$ data are shown as $\dndeta$, while
the
$\epem$ data (with cuts applied to reject events with large
initial-state photon radiation)
is shown as $dN/d{y_T}$, the rapidity distribution along the
event thrust axis, calculated assuming the pion mass.
The shapes of Au+Au and $\epem$ are similar (within 10\%) 
in shape and magnitude, especially within $|\eta|<4$.
In the lower panel, we also observe
that the shapes of the Au+Au and $\pbarp$ distributions
are also very similar over
a large range in $\eta$, but the integrals differ by $\sim40\%$.

\begin{figure}[h]
\begin{center}
\includegraphics[height=70mm]{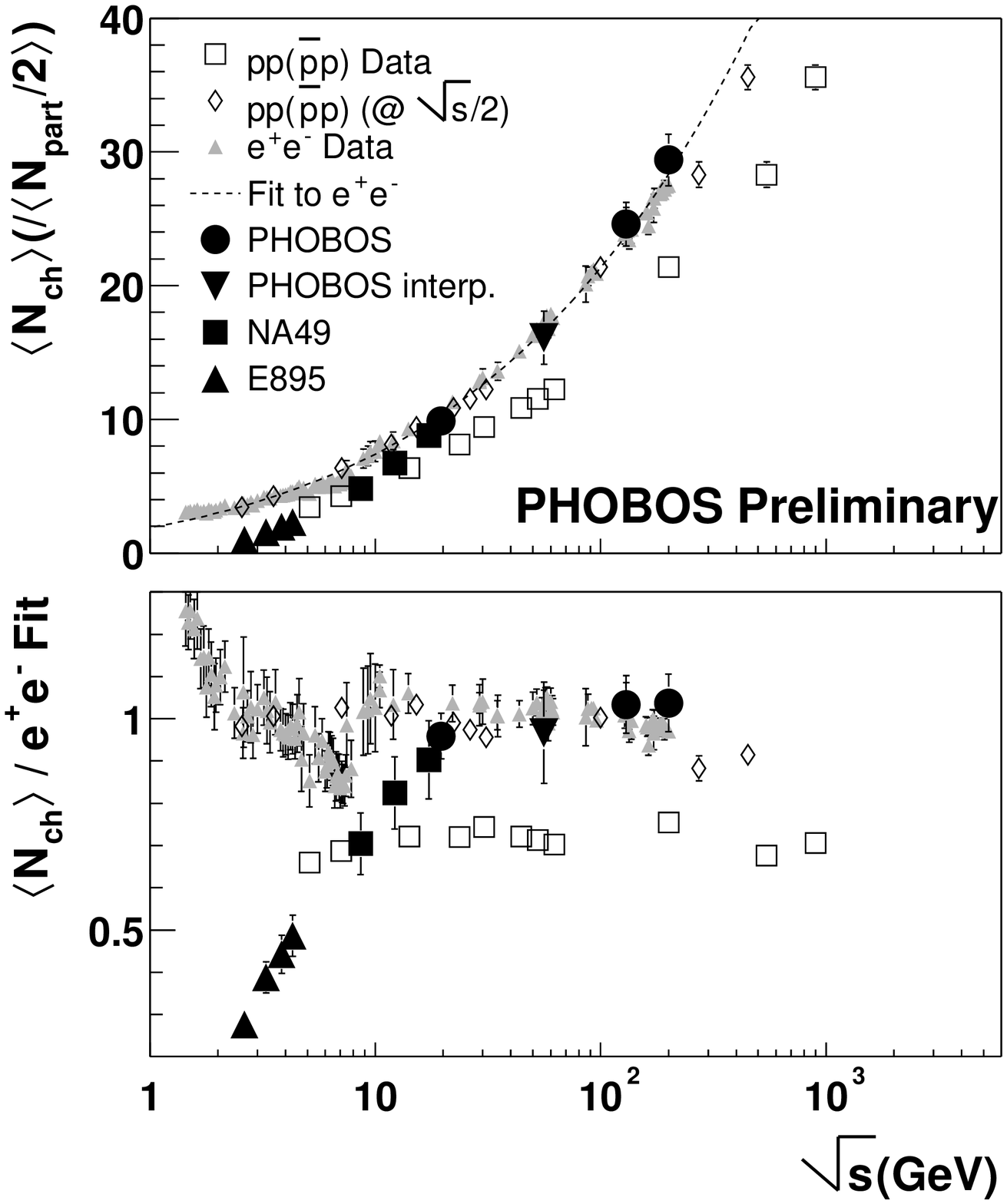}
\caption{
Comparison of $\ratio$ for A+A, $\pp/\pbarp$, and $\epem$
data compared with a fit to the $\epem$ data.
\label{fig:total_ratio}
}
\end{center}
\vspace*{-.8cm}
\end{figure}

In Fig. \ref{fig:total_ratio}, we compare $\ratio$ in heavy ion 
collisions \cite{heavy-ion} 
to $\epem$ and $\pp/\pbarp$ data over a large range in $\s$
\cite{Groom:in}.
It is observed that $\ratio$ lies below $\pp$ at low energies,
passes through the $\pp$ data around $\s\sim 10$ GeV, and then 
gradually joins with the $\epem$ trend above CERN SPS energies.
These comparisons can be seen more clearly by dividing all of the
data by a fit to the $\epem$ data \cite{Mueller:cq}.

The $\pp/\pbarp$ data follows the same trend as $\epem$, but it can be
shown that it matches very well if the ``effective energy'' $\seff = \s/2$
is used, which accounts for the leading particle effect seen
in $\pp$ collisions \cite{basile}.  
Ref. \cite{basile} finds that bulk particle
production in $\pp$ and $\epem$ data
does not depend in detail on the collision system but 
rather the energy available for particle production.
In this scenario, the Au+Au data suggests a substantially reduced leading
particle effect in central collisions of heavy nuclei at high energy.

\begin{figure}[h]
\begin{center}
\includegraphics[height=53mm]{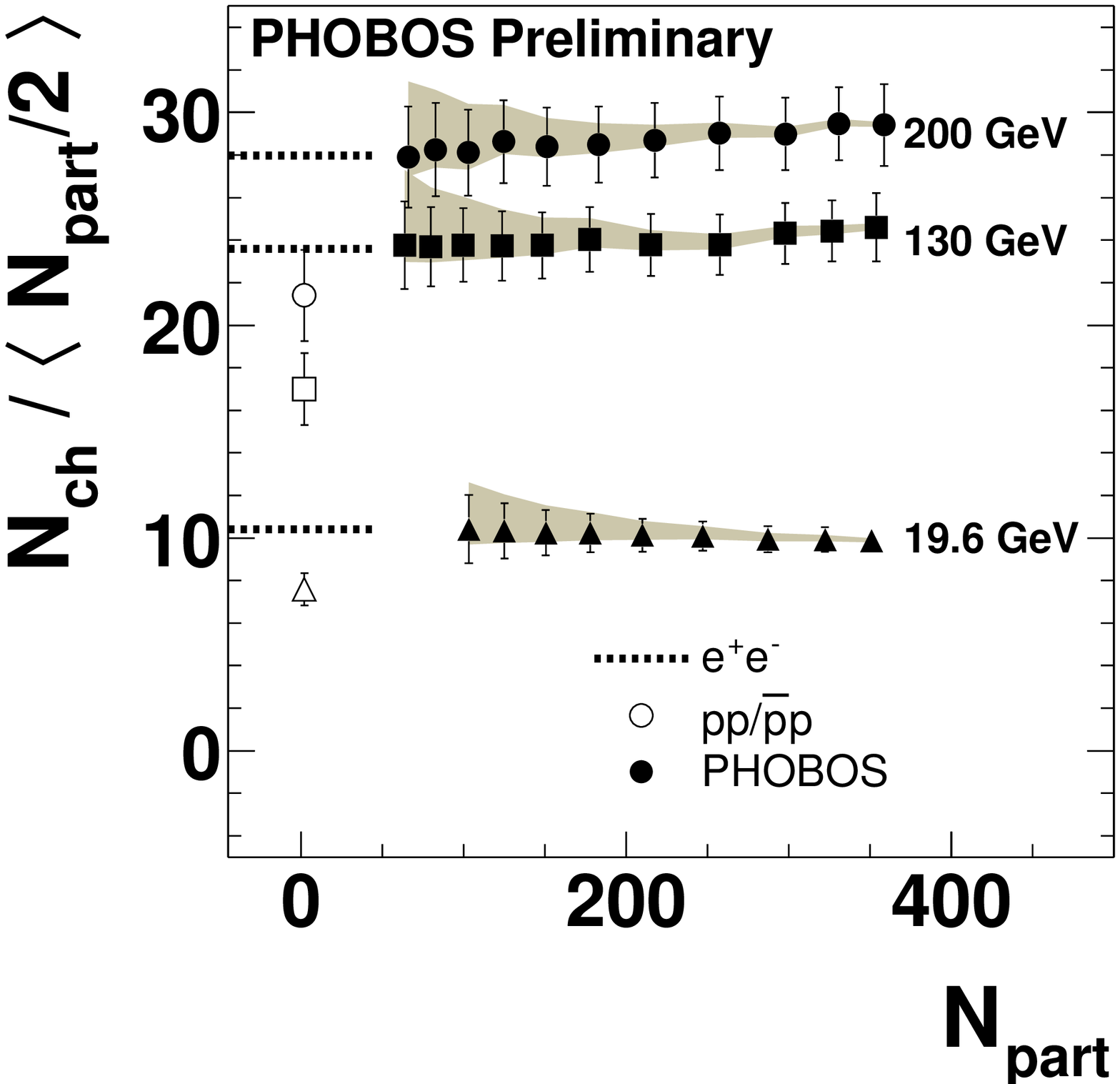}
\caption{
$\ratio$ vs. $\np$ for $\snn = $ 19.6, 130, and 200 GeV.
\label{fig:ntot}
}
\end{center}
\vspace*{-.5cm}
\end{figure}

The alleviation of the leading particle effect 
might not be so surprising in nuclear collisions.
Each participating nucleon is typically struck $\nubar = \frac{\nc}{\np/2}> 3$
times on average as it passes through the oncoming gold nucleus
for $\np>65$.  One could speculate that the multiple
collisions transfer much more of the initial longitudinal energy into
particle production.
This naturally leads to the scaling of total particle production
in heavy ion collisions with $\np$, as seen in Fig. \ref{fig:ntot},
reminiscent of the ``wounded nucleon model'' \cite{wounded}
but with the scaling factor determined by $\epem$ rather than $\pp$.

In conclusion, PHOBOS has observed two kinds of universal behavior. 
The first is an energy-independent, but centrality-dependent, 
universal limiting distribution
of charged particle production away from midrapidity.
The second is that the total charged particle multiplicity
per participant pair in heavy ion collisions above CERN SPS
energies scales with sqrt(s) in a similar way as $\epem$ collisions.
%
These 
features may offer a new perspective
on particle production in heavy ion collisions.

{\footnotesize
Special thanks to the ICHEP organizers and Heavy Ion convenors for the
invitation to speak at the conference.
This work was partially supported by U.S. DOE grants DE-AC02-98CH10886,
DE-FG02-93ER40802, DE-FC02-94ER40818, DE-FG02-94ER40865, DE-FG02-99ER41099, and
W-31-109-ENG-38 as well as NSF grants 9603486, 9722606 and 0072204.  The Polish
groups were partially supported by KBN grant 2 PO3B 10323.  The NCU group was
partially supported by NSC of Taiwan under contract NSC 89-2112-M-008-024.
}


\begin{thebibliography}{9}
\bibitem{phobos_limfrag} B.~B.~Back {\it et al.}, to be submitted to PRL.
\bibitem{ua5} G.~J.~Alner {\it et al.}, Z.\ Phys.\ C {\bf 33}, 1 (1986).
\bibitem{delphi} P.~Abreu {\it et al.}, Phys.\ Lett.\ B {\bf 459}, 397 (1999).
\bibitem{ALEPH} H. Stenzel, ALEPH Collaboration, Contributed paper to ICHEP2000 (2000).
\bibitem{heavy-ion} J. Klay, U.C. Davis PhD. Thesis (2001).  S.~V.~Afanasiev {\it et al.}, arXiv:nucl-ex/0205002 (2002).
\bibitem{Groom:in} D.~E.~Groom {\it et al.}, Eur.\ Phys.\ J.\ C {\bf 15}, 1 (2000).  
\bibitem{Mueller:cq} A.~H.~Mueller, Nucl.\ Phys.\ B {\bf 213}, 85 (1983).
\bibitem{basile} M.~Basile {\it et al.}, Phys.\ Lett.\ B {\bf 92}, 367 (1980).  M.~Basile {\it et al.}, Phys.\ Lett.\ B {\bf 95}, 311 (1980).
\bibitem{wounded} J.~E.~Elias {\it et al.} Phys.\ Rev.\ Lett.\  {\bf 41}, 285 (1978).  A.~Bia\l as, B.~Bleszy\'{n}ski and W.~Czy\.{z}, Nucl.\ Phys.\ {\bf B111}, 461 (1976).
\end{thebibliography}
\end{document}